\documentclass[journal]{IEEEtran}
\usepackage[utf8]{inputenc}
\usepackage[left=0.58in,right=0.58in,top=1.15in,bottom=0.8in]{geometry}
\usepackage{cite}
\usepackage{amsmath,amssymb,amsfonts}
\usepackage{graphicx}
\usepackage{subfigure}
\usepackage{textcomp}
\usepackage{xcolor}
\usepackage{algorithm}
\usepackage{algpseudocode}
\usepackage{bm}
\usepackage{hyperref}
\usepackage{url}
\usepackage{booktabs}

\newcommand{\revise}[1]{\textcolor{blue}{{\bf}  #1}}

\usepackage[compact]{titlesec}         
\titlespacing{\section}{5pt}{5pt}{5pt} 
\AtBeginDocument{
  \setlength\abovedisplayskip{5pt}
  \setlength\belowdisplayskip{5pt}}

\title{Carbon-Aware EV Charging}
\author{\IEEEauthorblockN{Kai-Wen Cheng\IEEEauthorrefmark{1}\IEEEauthorrefmark{4},Yuexin Bian\IEEEauthorrefmark{1}\IEEEauthorrefmark{4}, Yuanyuan Shi\IEEEauthorrefmark{1}, Yize Chen\IEEEauthorrefmark{2}}\\
\IEEEauthorblockA{\IEEEauthorrefmark{1} Department of Electrical and Computer Engineering,
University of California San Diego, San Diego, CA 92161
 }\\
\IEEEauthorblockA{\IEEEauthorrefmark{2}Lawrence Berkeley National Laboratory, Berkeley, CA 94720}\\

kac003@ucsd.edu, yubian@ucsd.edu, yyshi@eng.ucsd.edu, yizechen@lbl.gov

\footnote{\IEEEauthorrefmark{5} Authors contributed equally}}

\begin{document}

\maketitle
\begingroup\renewcommand\thefootnote{\textsection}
\footnotetext{Equal contribution}
\endgroup

\maketitle
\IEEEoverridecommandlockouts

\IEEEpubid{\makebox[\columnwidth]{978-1-6654-3254-2/22/\$31.00~\copyright2022 IEEE \hfill}\hspace{\columnsep}\makebox[\columnwidth]{ }}
\IEEEpubidadjcol

\begin{abstract}
    This paper examines the problem of optimizing the charging pattern of electric vehicles (EV) by taking real-time electricity grid carbon intensity into consideration. The objective of the proposed charging scheme is to minimize the carbon emissions contributed by EV charging events, while simultaneously satisfying constraints posed by EV user's  charging schedules, charging station transformer limits, and battery physical constraints. Using real-world EV charging data and California electricity generation records, this paper shows that our carbon-aware real-time charging scheme saves an average of $3.81\%$ of carbon emission while delivering satisfactory amount of energy. Furthermore, by using an adaptive balanced factor, we can reduce $26.00\%$ of carbon emission on average while compromising $12.61\%$ of total energy delivered.
\end{abstract}

\section{Introduction}
The adoption of electric vehicles (EV) is significant to the carbon neutrality goal since it helps reducing carbon emission of the transportation sector. Despite the numerous benefits of EV on the emission front, the rapid increase in EV charging events has become an emerging operational concern on the power grid level. With more than a quarter million of EVs sold in the United States in 2021 and more projected in the future, the grid-level power consumption will face an expedited increase. Moreover, since the transportation sector and the electric power accounts for $30\%$ and $25\%$ of carbon emissions~\cite{EPAReport} respectively, there are growing interests on how we can efficiently reduce grid-level carbon emissions in accompany with transportation electrification. While driving EVs decreases the direct carbon emission in transportation, charging EVs could possibly increase carbon emission in electricity generation due to mismatch in low carbon emission interval and peak charging interval. A German study indicated that if EV adoption rate exceeds $25\%$, we could see an increase in peak power demand of $30\%$ in some location ~\cite{mckinsey}. As the electricity generation energy mix becomes time-varying due to wind and photovoltaic (PV) energy source, increase in EV charging demand may lead to increasing use of fossil fuel generations during peak hours, which contradicts the carbon reduction goals. 

In this paper, we examine the possibility of designing carbon-intensity-adaptive charging policies to optimize the carbon trajectories of EV charging sessions. Take California as an example, the fast adoption of solar PV in California has made phenomenal impact on the state's grid energy mix, leading to the ``duck curve'' - the imbalance between electricity demand and renewable generation~\cite{duckcurve}. With respect to carbon emission, the ``duck curve'' also applies. In day time, the large amount of clean energy supply help lower the average carbon intensity. Depending on the season, we can also see a difference of 0.12 to 0.15 kgCO2/kWh between the day-time and night-time carbon intensity. This implies the potential of shifting EV charging demand to the hours with higher penetration of renewables~(low carbon intensity), therefore, absorbing the excess supply of electricity and reducing the overall carbon emission. 



\begin{figure}[t]
    \centering
    \includegraphics[width=0.9\columnwidth]{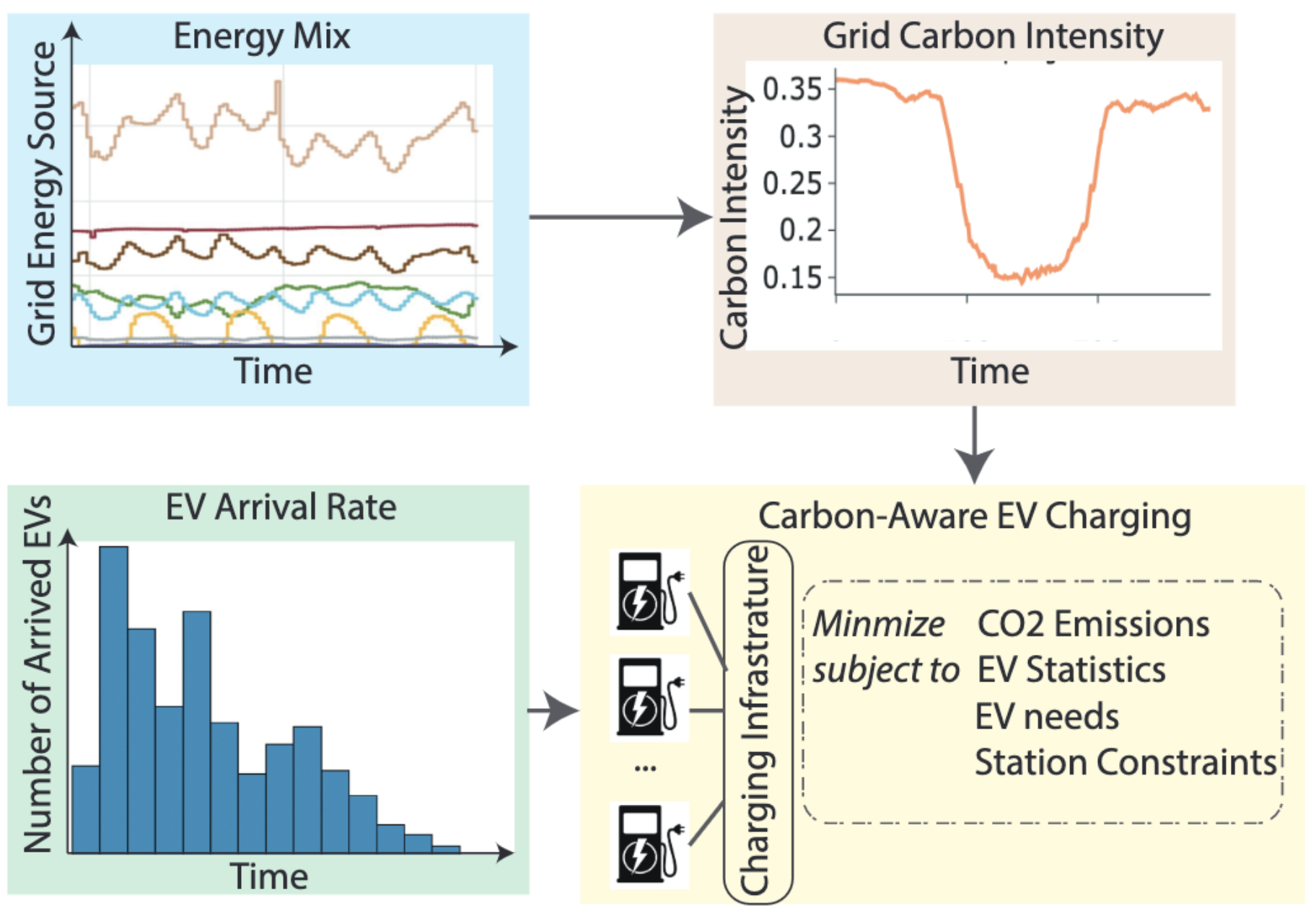}
    \caption{Schematic of proposed carbon-aware EV charging scheme. By investigating the flexibiliy coming from both grid carbon intensity and EV arrival/charging patterns, we are able to design adaptive charging algorithm to minimize charging station's carbon emissions. 
    The top left figure represents the time-series of the mix of different energy sources, which is detailed in Section III.A.}
    \label{fig:charging_diagram}
\end{figure}

While the flexibility in EV charging schedule can reduce the environmental impact, it is important to take individual charging session's time and energy constraint into consideration in real life.
In this paper, we take the perspective of an EV charging station operator that aim to reduce overall EV charging emission while satisfying individual user's charging needs. The goal is achieved by exploring the interactions among time-varying grid carbon intensity, EV charging patterns, and charging infrastructure's physical constraints.
Specifically, we would like to answer the following research question: \emph{what is the potential for reducing carbon emission of charging events by leveraging the flexibility of EV charging sessions?}

The contributions of this work are as follows. First of all, this paper introduces the carbon optimization problem in contrast to cost optimization for EV charging stations. With ongoing demand for low-carbon initiatives and action terms, this paper also brings carbon-awareness to the attention. On top of that, this paper shows the effectiveness of such optimization algorithm by solving the optimization with real-world data collected in the Berkeley Lab campus. Last, the introduction of carbon intensity forecasting enables online EV scheduling, which makes EV carbon minimization applicable in real-life using the the algorithm introduced\footnote{In this work, we consider the carbon emissions from charging sessions, excluding from, e.g., vehicle manufacturing,
 power unit construction or end-of-life.}.
 
 
 Numerical studies on real-world charging data from Berkeley campus and CAISO generation fuel mix data validate the effectiveness of the proposed method on optimizing EV charging events' carbon footprints in a real-time fashion. Our results show that the proposed charging scheme can save an average of 3.81$\%$ of carbon emission without sacrificing total delivered energy. 
 From our data and simulation, the charging station emits on average around 57 kgCO2 each day while delivering 288 kWh of energy. A 3.81$\%$ carbon reduction is equivalent to 792.7 kgCO2 per year. Moreover, with a 12.61$\%$ sacrifice on the energy delivery, we can further reduce the carbon emission by 26.00$\%$, which is around 5,143 kgCO2 per year. According to the latest social cost of carbon (SCC) estimate by the US government, the current SCC is priced at $\$$50 per ton~\cite{socialcost}. This indicates that our algorithm can reduce social cost by $\$257$ with a slight comprise of energy delivery. 
Assuming one charging session per day per EV in California, this means given current renewables penetration, we can reduce SCC by 5.47 million dollars per year. The cost reduction amount will continue to increase as more EVs are adopted in California.

\subsection{Related Work}
Researchers have been discussing the potential of designing EV charging scheduling algorithms to alleviate grid congestion~\cite{xiong2018vehicle}, to better integrate increasing renewables~\cite{gao2014integrated}, and to accommodate economical objectives as well as operating constraints~\cite{lee2018large}. Real-time algorithm considering arrival and departure statistics are discussed in \cite{tucker2022real}.
Bi-directional charging strategies and vehicle-to-grid (V2G) technologies have been also discussed in \cite{xu2020greenhouse}. 

There are some existing work on approaching load shifting problem via solving optimization problem explicitly \cite{chen2021smoothed,lee2019acn, quiros2015statistical} or via time-of-use pricing. Even though electricity prices follow some characteristic of the the load curve (lower cost during the noon and higher cost during the evening), cost optimization using TOU pricing is an indirect approach for optimizing carbon emissions. Based on a recent EV industry report~\cite{sense_report}, there is unutilized potential of carbon reduction via enabling flexible charging windows.

Carbon-aware EV charging scheduling is still novel in the way that it combines grid carbon intensity data, EV statistics and optimization framework. \cite{will2022consumer} conducted a survey on EV owners' preferences regarding carbon-neutral charging services. In \cite{ensslen2017empirical}, empirical studies validate that different geographical regions and charging schedules can lead to different carbon emissions.
\cite{colmenar2019electric} proposed a heuristic method to match EV charging with renewable energy resources on a macro scale without considering individual characteristics of charging sessions and charging station's physical constraint. In \cite{chen2019electric}, a demand response scheme with simplified charging models is proposed for the overall US market.

Our technique bears the closest resemblance to the industrial advances mentioned in \cite{parker2020,Sense2021}, and is partly inspired by the carbon-aware cloud computing considering flexible data center loads~\cite{radovanovic2021carbon}. The key distinction lies in how we formulate and solve the optimization problem by taking both carbon intensity forecasts and engineering constraints into account. We also develop carbon intensity forecasting method to accommodate real-time scheduling.



\section{Carbon-Aware Charging Framework}

In this paper, we consider the daily operation of an EV charging station. The charging station is connected to the electricity grid via a power distribution feeder. The charging current signal is communicated from the charging station to the vehicle's on-board charger. Upon arrival of each EV, the charging station is notified of their arrival battery energy level, requested energy and departure time according to inputs set by the user. 

We are interested in designing a charging scheduling algorithm to achieve station-level load shaping. In particular, since both the EV charging demand and grid-level carbon intensity are time-varying, there is potential to reduce station-level carbon emission by shifting charging load from periods with high carbon intensity to low carbon intensity periods, as long as the charging demand is satisfied before departure time.

The implemented algorithm is composed of several major components as are shown in Fig. \ref{fig:charging_diagram}: (i) extract and process grid-level energy mix data; (ii) predict the day-ahead carbon intensity (iii) receive and communicate session-specific charging statistics (arrival time, departure time, battery limits, and requested energy); and (iv) optimize carbon emissions with resource and charging schedule limits.

\section{Method}
\label{sec:model}
In this section, we firstly introduce carbon intensity calculation, and then illustrate how carbon reduction goal is integrated into charging station operation considering the stochasticity of EV charging requests and grid carbon intensity. 

\subsection{Carbon Intensity}
\begin{figure}[t]
    \centering
    \includegraphics[width=0.95\columnwidth]{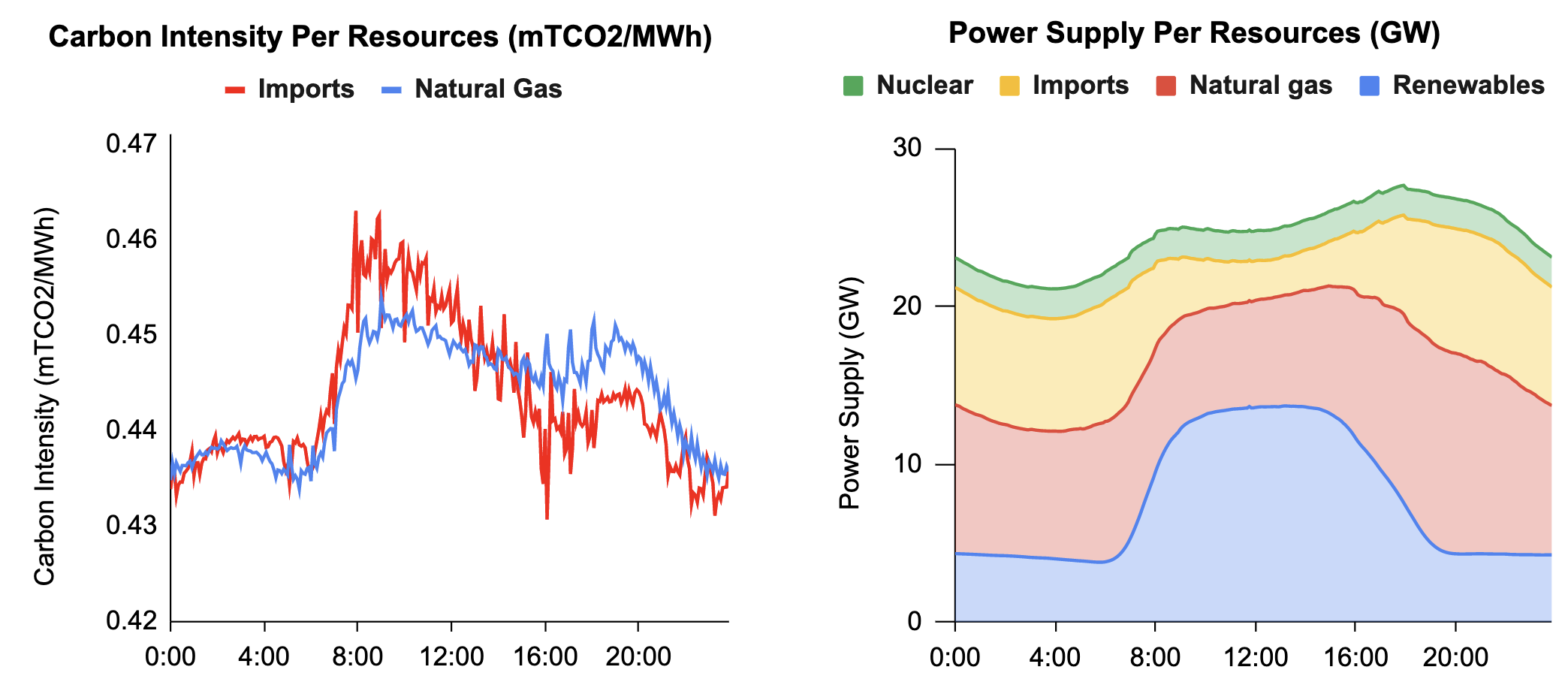}
    \caption{(Left) The carbon intensity of each resources. We exclude biogas, biomass, geothermal, and coal since it has relatively low carbon contribution in California. (Right) Daily average power supply per resource. We exclude coal and hydro since it is less significant. } 
    \label{fig:carbon_caiso}
\end{figure}

We denote the time-varying carbon emission (carbon intensity) as $C(t)$ with a unit of kgCO2/kWh or mTCO2(Metric ton CO2)/MWh. Carbon intensity data is a time-series data that reflects the grid-level carbon emission per energy unit. Carbon intensity depends on the power grid's operational plan and carbon intensity of each energy resource. 
In this paper, we specifically examined the carbon intensity for the California grid with data provided by CAISO (California Independent System Operator)\footnote{\url{http://www.caiso.com/todaysoutlook/pages/emissions.html}}. To calculate the carbon intensity, we consulted the carbon emission of the grid resources as well as the supply trend of the grid. Figure 2 describes the feature of the two data sets. In Fig \ref{fig:carbon_caiso} (left) there are two main contributors to the state's grid emission: imports and natural gas. 
\begin{equation}
    \label{equ:carbon intensity}
    C(t) = \sum_{i=1}^{N} C_i(t) \cdot \frac{P_i(t)}{P_{total}(t)}, \;  \forall i \in \mbox{grid mix};
\end{equation}
where $C_i(t)$ is a time-series data of the carbon intensity of each energy source, $P_i(t)$ is the power supplied by the energy source and $P_{total}(t)$ is the total power generated on the specific day at time $t$.
From the formula and the data, we can expect the carbon intensity to be high during the night and the early morning when solar energy is less available. On the other hand, the carbon intensity of the grid tends to be low during the day as solar energy and other renewables become a major contributor to the state's power supply.




\begin{figure}[H]
    \centering
    \includegraphics[width=0.75\columnwidth]{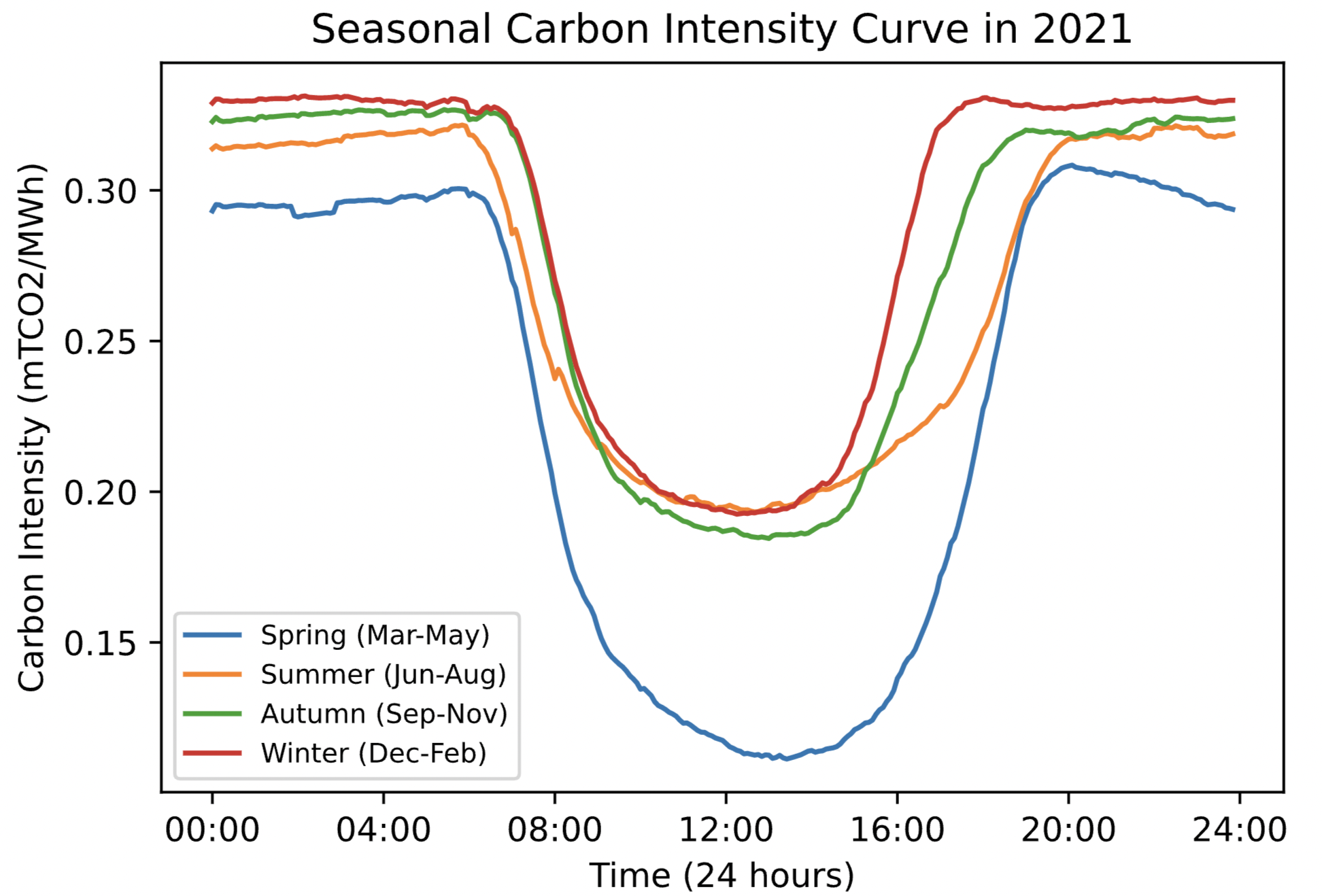}
    \caption{Average carbon intensity during different seasons.} 
\end{figure}
Carbon intensity is not only grid-mix dependent and time sensitive, but is also seasonally sensitive. Fig 3 shows an example of carbon intensity calculated using the above formula throughout the year. We notice that carbon intensity is lower especially during the spring. On top of that, we also notice a shift in time for the carbon intensity curve in autumn as well as winter. We can attribute this to the daylight saving hours policy in California. 

\subsection{Charging Station Operation Model}
As mentioned in Section II, our charging framework takes into account of the user constraints as well as the infrastructure constraints. These constraints are discussed below.
We use the discrete time model, where the time step is indexed in $\mathcal{T} := \{1,2,3,\ldots,T\}$.  $\mathcal{V}$ is the set of all EVs. For each vehicle $i \in \mathcal{V}$, we denote the charging demand information is $(x_{i,arrival}, x_{i,depart}, t_{i,arrival}, t_{i,depart})$, where $x_{i,arrival}$ is the state-of-charge SoC of EV $i$ upon its arrival of the charging station, $x_{i,depart}$ is the desired state-of-charge SOC of EV $i$ upon its departure. and $t_{i,arrival}, t_{i,depart}$ is the time index when the vehicle $i$ arrive in / depart from the charging station, $t_{i,arrival}, t_{i,depart} \in \mathcal{T}$.The above four factors are all externally determined by the user. Arrival time $t_{i,arrival}$ and initial energy level $x_{i,arrival}$ are determined when user plugged in the EV. Departure time $t_{i,depart}$ and charge $x_{i,depart}$ are input by the user when they initiate the charging session.

The model aims at deciding the charging power $u_i(t)$ for vehicle $i$ at time $t$, and the state-of-charge (SoC) $x_i(t)$ for EV $i$ at time $t$ given the specific objective. 
We are interested in solving the following optimization problem:
\begin{subequations}
    \label{equ:carbon_aware_EV}
     \begin{align}
    \min_{\mathbf{u}} \quad & \sum_{t \in \mathcal{T}} \sum_{i\in \mathcal{V}}  C(t)u_i(t) \Delta T + \lambda \sum_{i\in \mathcal{V}}|x_i(T)-x_{i,depart}| \label{equ:obj}\\
    s.t. \quad &  x_i(0)=x_{i,arrival}, \label{eq:initial_soc}\\ 
    &  x_{i}(t)=x_i(t-1)+ \frac{\delta u_i(t)}{E_i}, \; t\geq 1; \label{subequ:step}\\
        & 0\leq x_i(t) \leq \bar{x}_{i}, \; i\in \mathcal{V}; \label{subequ:soc}\\
    & 0\leq u_i(t) \leq \bar{u}_i, \; i\in \mathcal{V}; \label{subequ:action}\\
    & \sum_{i \in \mathcal{V}}  u_i(t) \leq \bar{P}, \quad t\in \mathcal{T};\label{subequ:total_u}\\
    & u_i(t)=0, \quad t \notin [t_{i,arrival},t_{i, depart}), \; i \in \mathcal{V}. \label{subequ:time}
     \end{align}
\end{subequations}
The above optimal charging scheduling problem minimizes the total carbon emissions while satisfying charging sessions' time and energy constraints, defined by the objective function~\eqref{equ:obj}. We firstly consider the case when $C(t)$ and EV charging demand information for next $T$ steps are available and communicated to the charging station operator. We denote $\Delta T$ to be the discretization interval, and in our case, $\Delta T$ is 5 minutes aligned with the carbon intensity measurement interval published by CAISO. $\lambda$ is a balanced factor to trade off between carbon reduction and energy delivery. The operational constraints considered in this framework are described as follows:
\newline \textbf{EV charging constraints}: Constraint \eqref{eq:initial_soc} defines the SoC level of EV upon arrival; and  \eqref{subequ:step} depicts the dynamics between charging power and battery SoC, where $E_i$ is the battery energy capacity (kWh). $\delta$ is an EV-specific charging parameter, which is known before the operation and is computed based on the charging efficiency and discretization, e.g., if charging efficiency is $90\%$, and discretization interval is 15 minutes, $\delta = 0.9 \times 15/60 = 0.225$. \eqref{subequ:soc} and \eqref{subequ:action} constrain the admissible charging power for single EV, where the battery maximum SoC is $\bar{x}_i$ and maximum charging power is $\bar{u}_i$. 
\newline \textbf{Charging station constraints}: \eqref{subequ:total_u} limits the total available charging power $\bar{P}$ at the charging station. We note that in this paper, we assume there is a known $\bar{P}$ decided by the charging station's transformer.
\newline \textbf{Temporal logic of charging sessions}: \eqref{subequ:time} describes the temporal logic that the charging session has to satisfy, as no charging power shall be delivered whenever EV $i$ has not arrived ($t < t_{i,arrival}$) or has departed $(t \geq t_{i,depart}).$
\newline\textbf{Other Factors to generalize}: The optimization problem that we introduced in (2a) is a general model that can be further generalized. For example, battery health is another important factor to be considered in EV charging. A more complex version of this optimization problem could potentially include penalty terms in the objective to discourage sudden change in charging power, which prevents battery health from deteriorating. 


\section{Real-Time Scheduling}
In this section, we propose a forecasting module to inform charging station operators day-ahead carbon intensity, and extend the framework described in Sec \ref{sec:model} to  real-time settings.

Offline algorithm assumes we have access to all the information including future arrival statistics. While in the real-time algorithm, we assume we can only decide charging rate based on the predicted carbon intensity and information of arrived electric vehicles.
An online algorithm is crucial for the real life applicability of our charging strategy because there are two main factors that changes daily in our optimization problem: daily average carbon intensity and daily EV arrival pattern. From figure 5 and 6, we can see that daily EV arrival varies from 5 vehicles to more than 40 vehicles per day. An online scheduling algorithm will be able to adjust scheduling the given arrival of EV's. On top of that, our discussion of carbon intensity also demonstrated seasonal volatility. With EV arrival uncertainty combined with carbon intensity uncertainty, and online algorithm allows the charging station to adjust it's charging strategy given different scenario. 

\subsection{Carbon Intensity Day-Ahead Forecasting}
\begin{figure}[t]
    \centering
    \includegraphics[width=0.8\columnwidth]{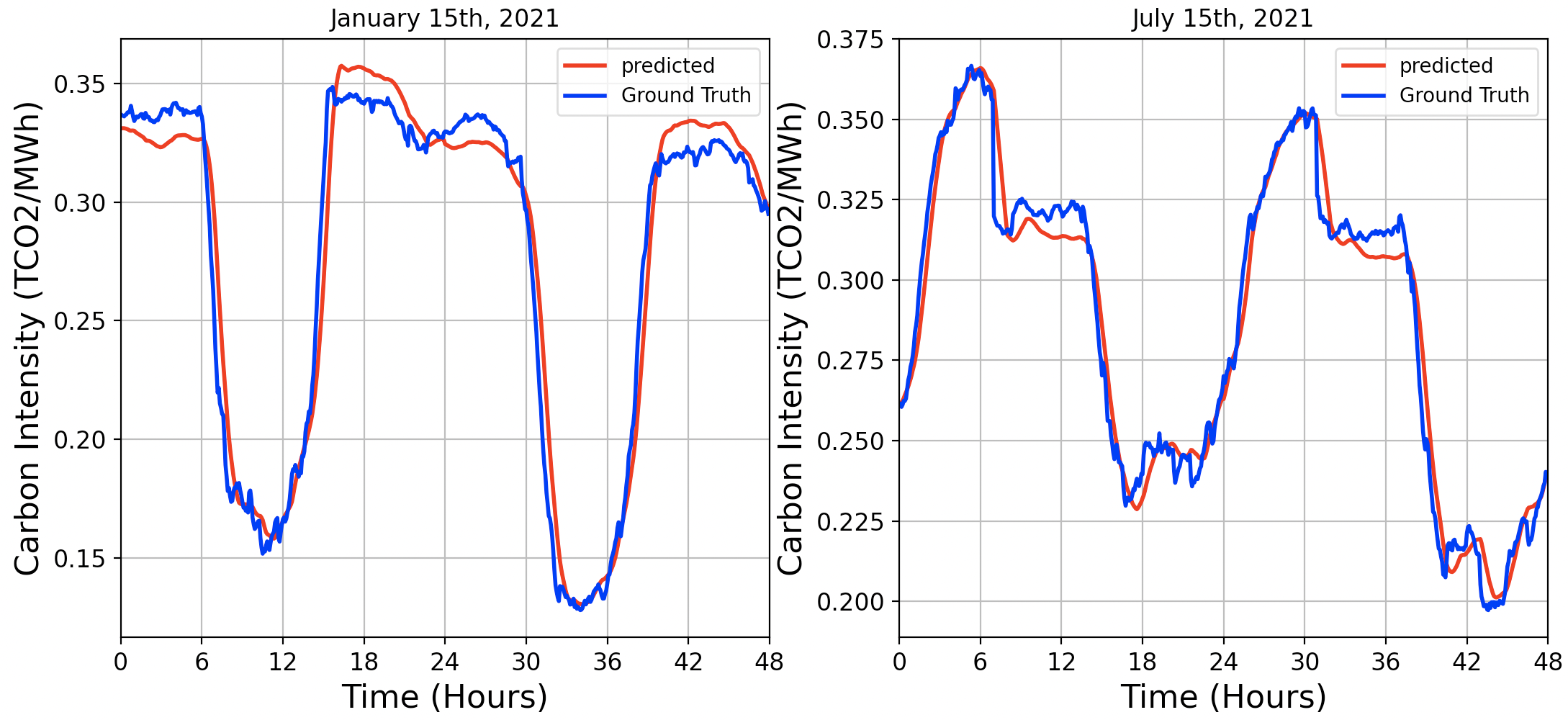}
    \caption{The ground truth (blue) and predicted carbon intensity (red) for two-day sample in January and July respectively.}
    \label{fig:carbon_forecast}
\end{figure}

We use the 2021 CAISO data (shown in Fig~\ref{fig:carbon_caiso}) and CAISO day-ahead load forecast data $Load(t)$ to fit and evaluate the carbon predictor model. The grid's day-ahead carbon intensity $\hat{C}(t)$ is closely related to the future electricity demand, calendar features, and past carbon intensity patterns. We randomly choose $80\%$ data to fit the linear regression model and rest data for testing the model performance. The resulting forecasting model is as follows: 
\begin{equation}
\begin{split}
\label{equ:carbon_predict}
    \hat{C}(t)&=\beta_0+\beta_1\times Minute + \beta_2 \times Hour + \beta_3 \times Day \\&+ \beta_4 \times Month + \beta_5 \times Load(t)+ \beta_6 \times MA(C(t))_{24}\\&+\beta_7 \times MA(C(t))_{12}+\beta_8 \times MA(C(t))_{1};
    \end{split}
\end{equation}
where $MA(\cdot)_{a}$ denotes the moving average of past $a$ hours. Fig~\ref{fig:carbon_forecast} illustrates the ground truth carbon intensity and forecasted carbon intensity using the fitted model \eqref{equ:carbon_predict}. We note that other statistical and machine learning models can also be used for carbon intensity prediction. We use the linear regression model since it is simple and provides accurate prediction result. For the CAISO case study, the overall testing mean squared absolute error is $0.0085$. 
We also demonstrate such forecasts can be readily integrated into the proposed charging scheduler without affecting the algorithm performance. Power grid CO2 emission intensity forecast is also investigated in \cite{leerbeck2020short} by considering additional features and an autoregressive model. 
We calculate and forecast the average carbon intensity rather than marginal carbon intensity, as in this study, we focus on the operating strategies for single EV charging station. Such single-station operation scheme shall pose minimal effects to transmission grid carbon profiles.  We refer to \cite{carbon2022online} for a more complete review and open-source data on electricity grid carbon intensities.

\subsection{Online Algorithm}
With the predicted carbon intensity $\hat{C}(t)$, we are ready to implement proposed carbon-aware charging scheduler in an online fashion. Denote the optimization problem \eqref{equ:carbon_aware_EV} as $\textbf{SCH}$ with input information coming from the charging infrastructure and EVs' peak power constraints $\bar{P}$ and $\bar{u}$, predicted carbon intensity, the active charging vehicles $\mathcal{V}_k$ along with their states $x_{\mathcal{V}_k}$ where $k \in \mathcal{K}, \mathcal{K} := \{1,2,3,\ldots,K\}$, the algorithm is summarized in Algorithm \ref{alg:algorithm}. 
\revise{}

\begin{algorithm}
\caption{Online Carbon Aware EV Charging}
\label{alg:algorithm}
\begin{algorithmic}
\Require Charging infrastructure and each EV's peak charging power limit $\bar{P}$ and $\bar{u}$, operation time length $K$. 
\Ensure Charging power $(u_i(1), u_i(2), \ldots, u_i(K), i \in \mathcal{V})$.
\For{$k = 1, \ldots, K$}
\State  $\mathcal{C}_k: = \{\hat{C}(k), \ldots, \hat{C}(k+T-1)\}$
\State  $\mathcal{V}_k :=  \{i \in \mathcal{V} | x_{i,arrival} \leq k \textbf{ AND }  x_{i,depart} > k\}$
\State  $x_{\mathcal{V}_k} :=  \{x_i(k) | i \in \mathcal{V}_k \}$
\State \textbf{compute} 
\State \quad
 $(u^*_i(1), \ldots, u^*_i(T), i \in \mathcal{V}_k) = \textbf{SCH}(\mathcal{C}_k, \mathcal{V}_k, x_{\mathcal{V}_k})$ 
\State \textbf{update} $u_i(k) = u^*_i(1)\,, \forall i \in \mathcal{V}_k$ and $u_i(k) = 0, i \not\in \mathcal{V}_k$.
\State \textbf{update} $x_i(k+1) = x_i(k) + \frac{\delta u_i(k)}{E_i}, \forall i \in \mathcal{V}$.
\EndFor
\end{algorithmic}
\end{algorithm}

At each time step $k = 1, 2, \ldots, K$, we take the forecasted carbon intensity for the next $T$ steps, $\mathcal{C}_k: = \{\hat{C}(k), \ldots, \hat{C}(k+T-1)\}$, the current available EVs $\mathcal{V}_k$ and their SoC levels $x_{\mathcal{V}_k}$, and solves the optimization~\eqref{equ:carbon_aware_EV}. With the optimized charging schedule $(u^*_i(1), \ldots, u^*_i(T), i \in V_k)$, we assign  $u^*_i(1)$ to the charging power of EV $i$ at time step $k$ to $u_i(k) = u^*_i(1)$ for the active vehicles and update state of charge accordingly for every vehicle $x_i(k+1) = x_i(k) + \frac{\delta u_i(k)}{E_i}, \forall i \in \mathcal{V}$.

\section{Simulation Results}
\label{sec:simulation}
We use real-world, campus-wide EV charging station data from the Lawrence Berkeley National Laboratory campus for the experiment, which contains all charging sessions from the year of 2021 including EV arrival time, energy consumption and depart time. We assume the battery capacity is 50 kWh for all the vehicles, the available charging power from the charging station is 180 kW, and maximum charging rate is 7.5 kW\footnote{We aggregate all on-campus EV chargers together as a single charging station and evaluate the algorithm.}. 
Without loss of generality, we choose the simulation interval $\Delta T$ as 5 minutes, and there are 288 time steps per day. Source code and data for our experiments are available at \url{https://github.com/kaicheng0824/carbon_aware_ev_charging}.
\begin{figure}[t]
    \centering
    \includegraphics[width=0.9\columnwidth]{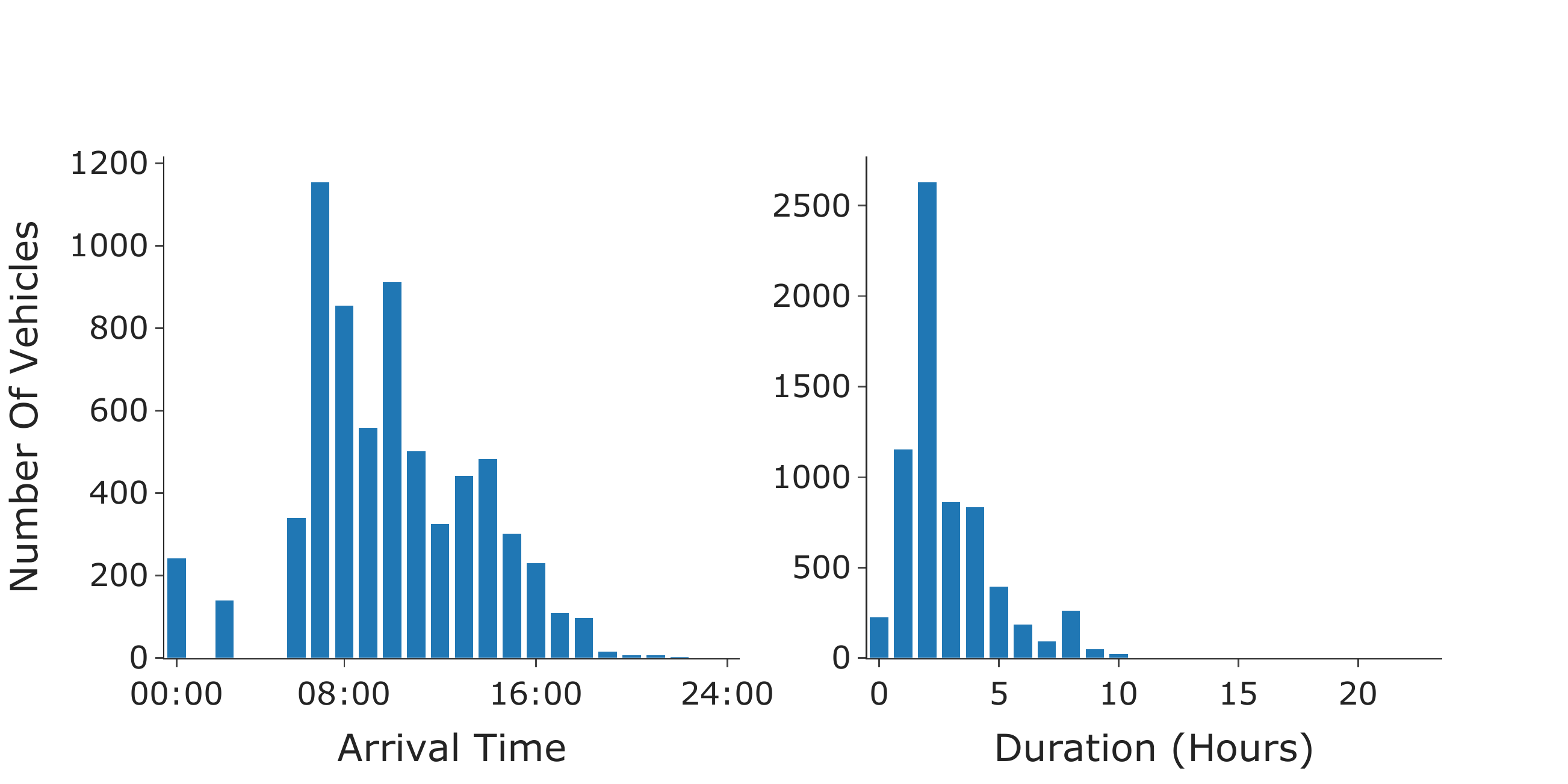}
    \caption{The distribution of arrival time (left) and duration (right) of EV charging sessions in 2021.}
    \label{fig:arrivalev}
\end{figure}

We include a brief introduction of the benchmarking EV charging scheduling methods:
\begin{itemize}
    \item The equal sharing (ES) algorithm allocates available charging power supply $\bar{P}$ \emph{equally} to all active EV charging sessions at each time step, while satisfying each EV’s maximum charging rate and energy constraint.
    \item The earliest deadline first (EDF) algorithm assigns the maximum allowable charging power $\bar{u}_i(t)$ to EVs with earliest departure time while keeping the sum of charging power less or equal to the infrastructure power limit $\bar{P}$.
    \item The time of use price aware (TOU) algorithm solves the optimization problem that minimizes the charging cost and required energy deviation.
\end{itemize}

We report energy delivery quality EDQ (\%) for the overall charging station (EDQ-Station) and the average energy delivery quality (\%) per session (EDQ-Session) for performance evaluation according to ~\eqref{eq:metric}: 
\begin{equation}\label{eq:metric}
\begin{aligned}
     \text{EDQ-Station} = \frac{\sum_{i \in \mathcal{V}}|x_{i,T}-x_{i,arrival}|}{\sum_{i \in \mathcal{V}}|x_{i,depart}-x_{i,arrival}|}, \\
     \text{EDQ-Session} = \frac{1}{|\mathcal{V}|}\sum_{i\in \mathcal{V}}\frac{|x_{i,T}-x_{i,arrival}|}{|x_{i,depart}-x_{i,arrival}|}.
\end{aligned}
\end{equation}


Fig~\ref{fig:arrivalev} shows the distribution of vehicle arrival time and yearly staying time. We can see that most vehicles arrive before noon when the carbon intensity is lower, which leave room for charging flexibility. Most arrived vehicles stay for less than four hours in the charging station. 

\begin{figure*}[t]
    \centering
    \includegraphics[width=1.67\columnwidth]{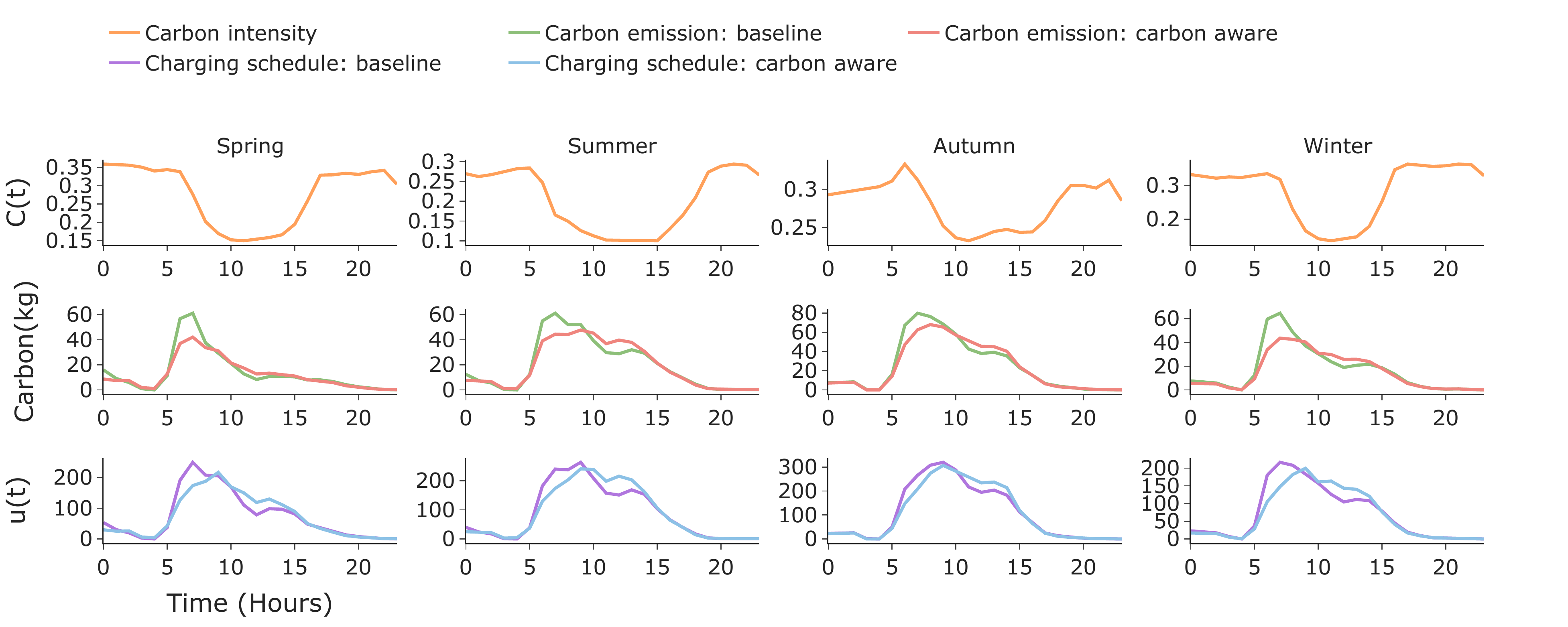}
    \caption{The carbon emission shift compared with the baseline (EDF) in four seasons. Carbon intensity exhibits the feature of duck curve among four seasons, and our method takes advantage of the low carbon intensity pattern and tends to postpone the charging to the time when carbon intensity becomes lower.}
    \label{fig:season}
\end{figure*}

\subsubsection{Seasonal Shift}
We first present the seasonal carbon emission shift using our method and the baseline EDF method. We select days from March 1 to May 31 as spring, June 1 to August 31 as summer, September 1 to November 30 as autumn and December 1 to February 28 as winter. Fig~\ref{fig:season} shows the comparison of carbon emission and energy delivery (\%) among four seasons with the balanced factor $\lambda = 0.35$. 
From the figure, we can see that carbon intensity is lower at noon than other time. Our method takes advantage of the low carbon intensity pattern and shifts the flexible charging demand to the time when carbon intensity is low, such that the carbon emission is reduced efficiently.

\subsubsection{The impact of the balanced factor}\label{sec:impact}
\begin{figure}[h]
    \centering
    \includegraphics[width=0.8\columnwidth]{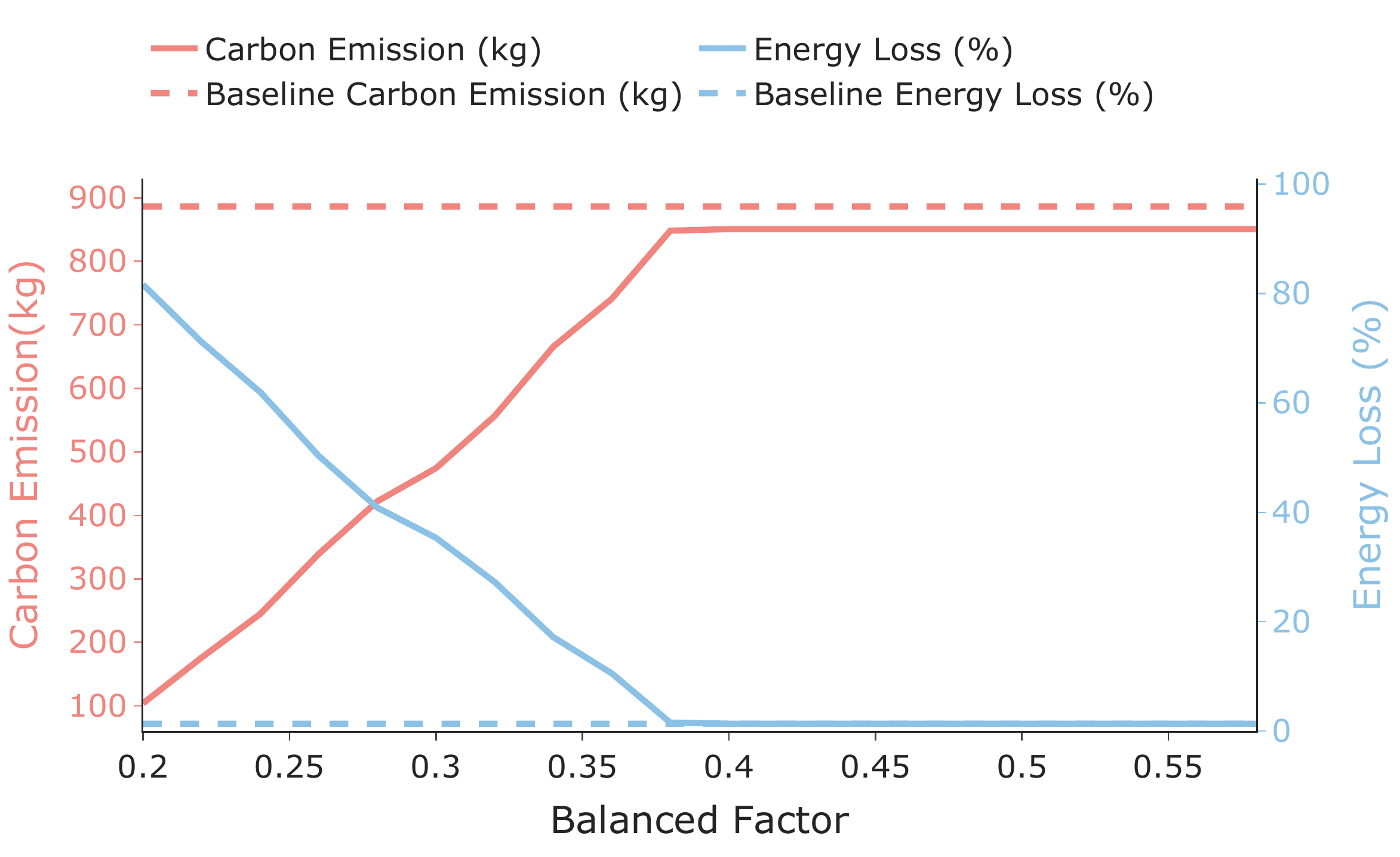}
    \caption{The tradeoff between the total carbon emission and percentages of energy loss among different balanced factor using our method and baseline (EDF). We use the first 20 days to simulate the result.}
    \label{fig:diff_factor}
\end{figure}

We use the first 20 days in the dataset to show how balanced factor $\lambda$ in ~\eqref{equ:obj} influences the carbon emission and energy delivery quality. Fig~\ref{fig:diff_factor} shows the change of total carbon emission (red curve) and energy loss (blue curve) using different balanced factors. We can see that with the balanced factor decreasing, the carbon emission will decrease while the energy loss will increase. We also compare the performance with the baseline EDF method. Using a balanced factor $\lambda \geq 0.4$, we can achieve the {same} energy delivery while reducing 5.6\% CO2 for 20 days.


\subsubsection{Comparison with baselines}
\begin{figure}[htbp]
    \centering
    \includegraphics[width=0.8\columnwidth]{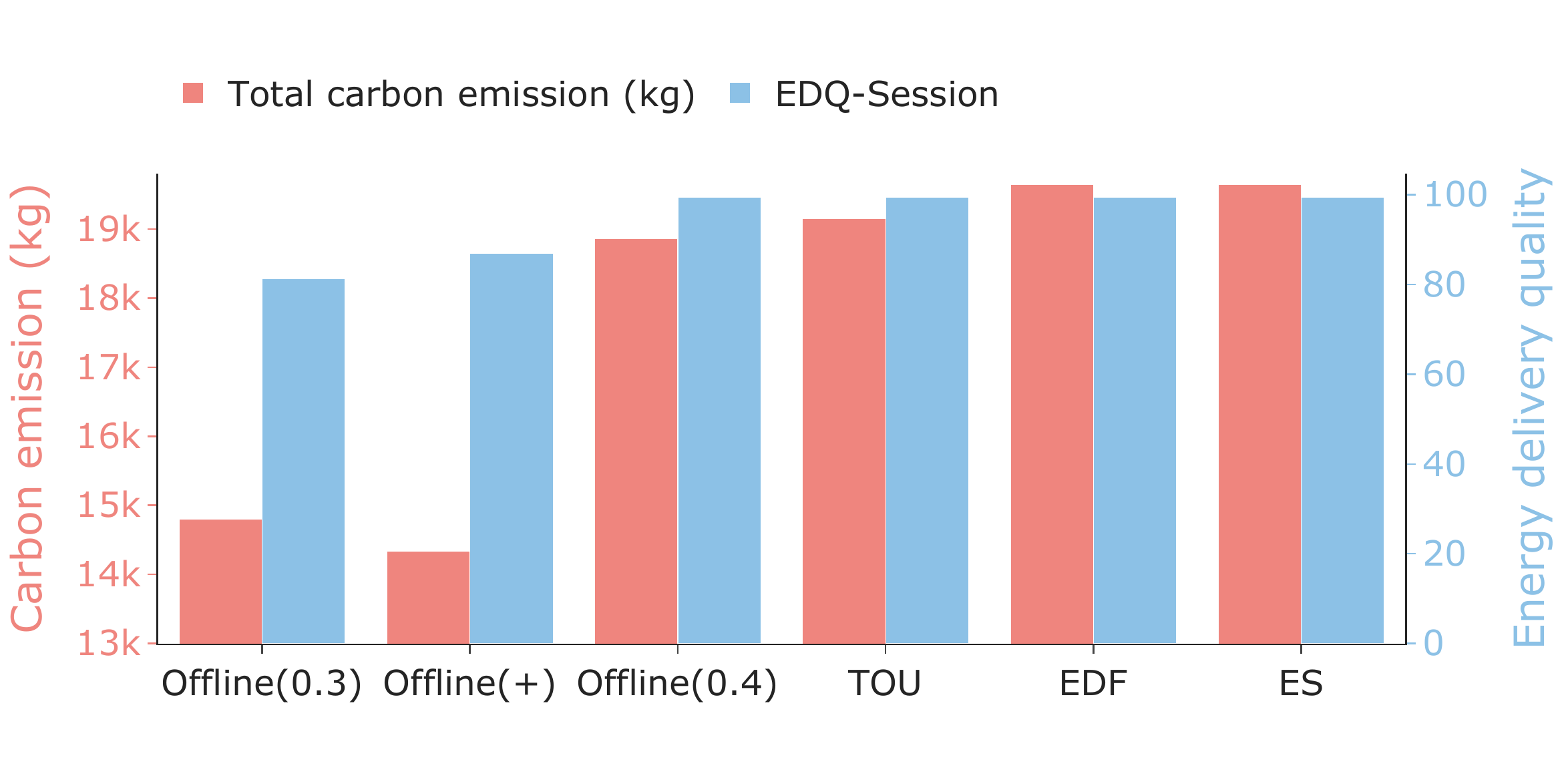}
    \caption{The comparison of average energy deliver quality per session (EDQ-Session) and carbon emissions: offline (0.3), offline (+) (smarter balanced factor strategy), offline (0.4), ToU aware (TOU), earliest deadline first (EDF), equal sharing (ES). }
    \label{fig:compare_adapted}
\end{figure}

Here we use the entire dataset in 2021. We compare our method with three baseline methods: ES, EDF and TOU. 
We test our method with a fix balanced factor $\lambda = 0.3$ (denoted as \texttt{Offline(0.3)}) and $\lambda = 0.4$ (denoted as \texttt{Offline(0.4)}), respectively. We also test the carbon-aware charging algorithm with a flexible factor scheme (denoted as \texttt{Offline(+)}), where we choose $\lambda = 0.3$ for days with vehicle size $n \leq 20$, $\lambda = 0.25$ for days with $n \in [21,30]$ and $\lambda = 0.35$ for other cases. 
Secion~\ref{sec:impact} shows that larger balanced factor is more likely to deliver the required energy while leading to less carbon reduction. Since group one (less than 20 charging sessions per day) has relatively low carbon emission, and group three (more than 30 sessions charging sessions per day) has less flexibility in shifting charging demand, we choose larger balanced factor for group one and three to prioritize the energy delivery.
The results are shown in Figure~\ref{fig:compare_adapted}. With $\lambda = 0.4$, the offline optimization can reduce $3.97\%$ carbon emission while slightly improves the energy delivery quality by $0.03\%$ compared to EDF and ES, and it can reduce $1.50\%$ carbon emission and maintain the same energy delivery quality compared to TOU. This is because EDF and ES will schedule charging instantly once EVs arrive, while TOU and our approach adjust the charging power by solving the current optimization instance. TOU rate and carbon intensity is higher after 4pm. To minimize the energy cost, ToU-aware policy is likely to encourage chargers to finish charging before 4pm, and in this way it can also partially help reduce the carbon emission. 
With $\lambda = 0.3$, the carbon aware optimization method \texttt{Offline(0.3)} can save $24.7\%$ and $22.71\%$ carbon emission compared to EDF/ES and TOU respectively, 
while also lead to $18.20\%$ in reduction in energy delivery quality.  
With flexible balanced factors, \texttt{Offline(+)} saves $3.17\%$ carbon and improves $5.55\%$ in delivery quality compared to \texttt{Offline(0.3)}. An interesting future direction is to develop a more principled method of adaptively choosing the balanced factor according to daily charging patterns, that can both maintain the high energy delivery quality and effectively reduce carbon emission.

\begin{table}[htbp]
    \centering
    \begin{tabular}{c c}
    \hline 
      \textbf{Strategy}   &  \textbf{Carbon emission (kg)} \\
      \hline
      Equal Sharing   &  2.913 \\
      Earliest Deadline First & 2.913 \\
      ToU price aware &  2.840 \\
      Offline (0.4) & 2.798 \\
      Offline (+)   & 2.127 \\
      Offline (0.3) & 2.195 \\
      \hline
    \end{tabular}
    \caption{Average carbon emission for each charging session using different approaches.}
    \label{tab:compare}
\end{table}
Table~\ref{tab:compare} shows the average carbon emission for each charging session. Compared to EDF, offline with $\lambda=0.4$ reduces 0.115 kg CO2, which is equivalent to travelling 1.15 miles\footnote{\href{https://www.tesla.com/ns_videos/2021-tesla-impact-report.pdf}{Tesla’s 2021 Impact Report}}; offline with adapted balanced factors saves 0.786 kg CO2, corresponding to the travelling distance of 7.86 miles.

\subsection{Online EV Charging Scheduling Results}
We next demonstrate the performance of the online scheduling algorithm. We use the carbon intensity prediction and online EV information to generate the charging schedule, that is, at each time step $t$, the model only knows the information of arrived vehicles, and use predicted carbon intensity data for the optimization.

\begin{figure}[h]
    \centering
    \includegraphics[width=0.8\columnwidth]{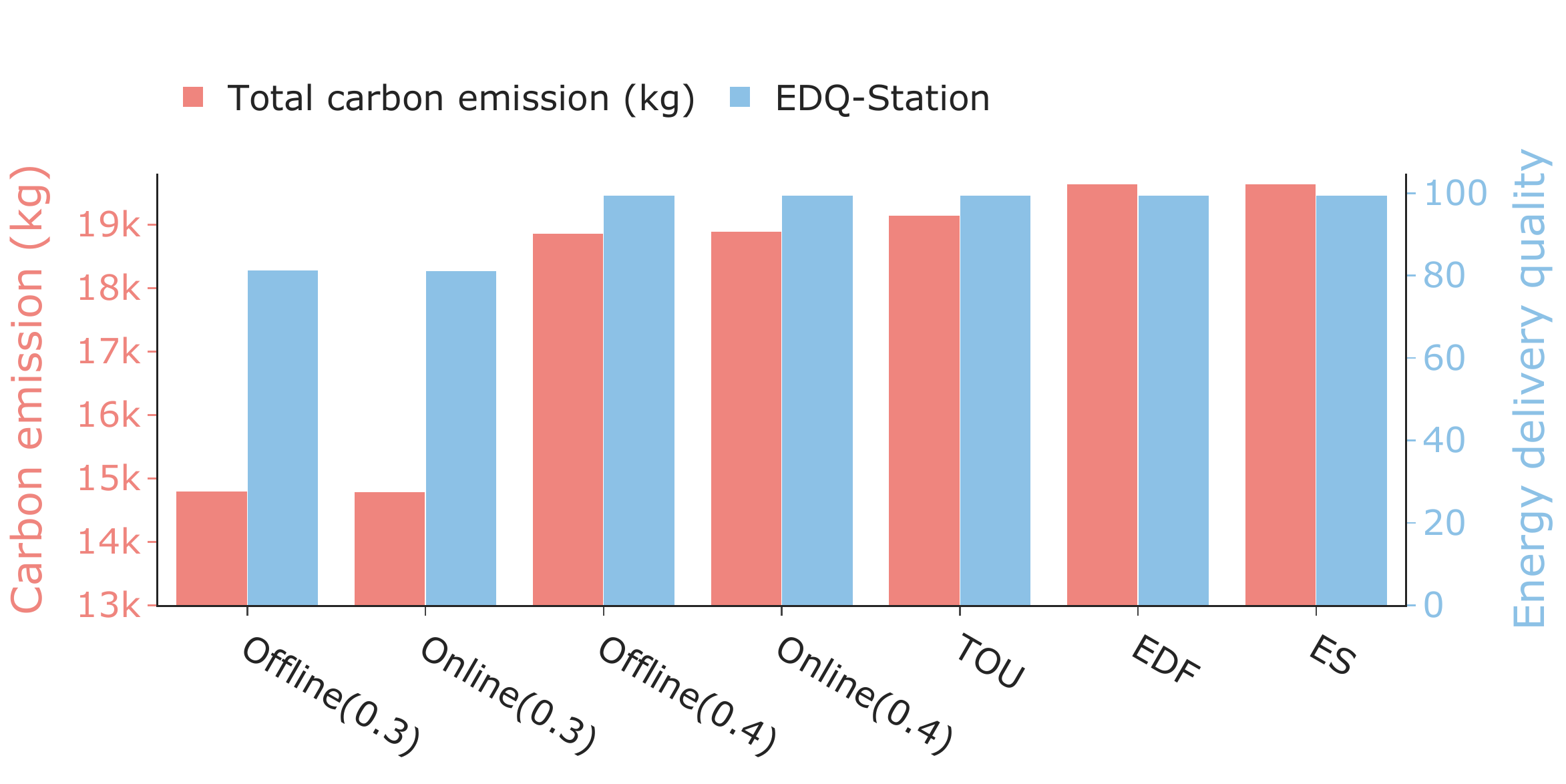}
    \caption{The comparison of energy delivery quality for station (EDQ-Station) and carbon emission: offline (carbon aware offline), online (real time carbon aware), ToU aware (TOU), EDF, and ES. Proposed online algorithm achieves similar performances compared to offline benchmarks.}
    \label{fig:online}
\end{figure}

Fig~\ref{fig:online} shows that our online algorithm does not hurt the performance much compared to the offline optima. With $\lambda = 0.3$, the online algorithm delivers $0.21\%$ less energy and generates $0.08\%$ less carbon emission compared to the offline algorithm.
With $\lambda = 0.4$, the online algorithm delivers $0.002\%$ less energy and generates $0.15\%$ more carbon emission compared to the offline algorithm.
Compared to EDF and ES, the real-time scheduling saves $3.81\%$ of carbon emission and can keep delivering the same energy; compared to TOU, it saves $1.50\%$ carbon and maintain the same energy delivery quality.
With balanced factor $\lambda = 0.3$, the online algorithm can further reduce $27.01\%$ of carbon emission at the cost of delivering $14.21\%$ less energy than EDF and ES; and reduces $25.13\%$ carbon at the cost of delivering $12.61\%$ less energy compared to TOU. All results indicate the online algorithm is highly efficient compared to the offline solutions.

\section{Conclusion and Future Work}
In this paper, we explore how EV charging scheduling will positively affect the carbon emissions of EV charging. More interestingly, in this paper, we use the grid-level average carbon intensity as an indicator for optimizing charging facilities' emission profiles. In the future work, we will investigate the choices of carbon and economical objectives, and explore the planning and operation mechanisms considering multiple EV charging stations. We will also explore how pricing mechanisms can affect the session-specific carbon emissions.

\bibliographystyle{IEEEtran}
\bibliography{bib}
\end{document}